\documentclass[useAMS,usenatbib]{mn2e}
\usepackage{graphicx}
\usepackage{amsmath}
\usepackage{ctable}

\newcommand{\kms}{\ensuremath{\mathrm{km\,s^{-1}}}}

\title[Dark-Matter Domination and a Salpeter-type IMF in a Massive
  Early-type Galaxy]{The X-Shooter Lens Survey - I. Dark-Matter
  Domination and a Salpeter-type IMF in a Massive Early-type Galaxy}

\author[]{C. Spiniello$^{1}$\thanks{E-mail:
spiniello@astro.rug.nl}, L.V.E. Koopmans$^{1}$, S.C. Trager$^{1}$, O. Czoske$^{2}$, T. Treu$^{3}$\\~\\
$^{1}$Kapteyn Institute, University of Groningen, P.O. Box 800, 9700 AV Groningen, the Netherlands\\
$^{2}$Institut  f\"{u}r Astronomie, Universit\"{a}t Wien,T\"{u}̈rkenschanzstra$\beta$e 17, 1180 Wien, Austria \\
$^{3}$Department of Physics, University of California Santa Barbara, Santa Barbara, USA} 

\begin{document}

\pagerange{\pageref{firstpage}--\pageref{lastpage}} \pubyear{2010}
\maketitle
\label{firstpage}

\begin{abstract}
We present the first results from the X-Shooter Lens Survey (XLENS):
an analysis of the massive early-type galaxy SDSS J1148+1930 at
redshift $z=0.444$.  We combine its extended kinematic profile --
derived from spectra obtained with X-Shooter on ESO {\sl Very
  Large Telescope} -- with strong gravitational lensing and
multi-color information derived from Sloan Digital Sky Survey (SDSS) images.
Our main results are (i) the luminosity-weighted stellar velocity
dispersion is $\langle \sigma_{*}\rangle(\la R_{\rm
  eff})=352\pm10\pm16\,\kms$, extracted from a rectangular aperture of $1\farcs8 \times 1\farcs6$
  centered on the galaxy, more accurate and considerably lower than a
previously published value of $\sim450\,\kms$; (ii) a single-component
(stellar plus dark) mass model of the lens galaxy yields a logarithmic
total-density slope of $\gamma'=1.72^{+0.05}_{-0.06}$ (68 per cent CL;
$\rho_{\rm tot} \propto r^{-\gamma'}$) within a projected radius of $\sim 2\farcs 16$; 
(iii) the projected stellar mass fraction, 
derived solely from the lensing and dynamical data, is
$f_{*}(<R_{\rm E}) = 0.19^{+0.04}_{-0.09}$~(68 per cent CL) inside the
Einstein radius for a Hernquist stellar profile and no anisotropy. The
dark-matter fraction inside the effective radius $f_{\rm DM}(<R_{\rm
  eff}) = 0.60^{+0.15}_{-0.06} \pm 0.1$ (68 per cent CL), where the
latter error is systematic; (iv) based on the SDSS colors, we find
$f_{*, \rm Salp}(<R_{\rm E})=0.17 \pm 0.06$ for a Salpeter Initial Mass Function (IMF) and
$f_{*,\rm Chab}(<R_{\rm E})=0.07 \pm 0.02$ for a Chabrier
IMF. 
The lensing and dynamics constraints on the stellar mass fraction agree well with those
independently derived from the SDSS colors for a Salpeter IMF, which is preferred over a Chabrier IMF at variance with standard results for lower mass galaxies.
Dwarf-rich IMFs in the lower mass range of
$0.1$--$0.7\,M_{\odot}$, with $\alpha\ge 3$ (with $dN/dM \propto
M^{-\alpha}$) -- such as that recently suggested for
massive early-type galaxies with $\alpha = 3$ in the mass range $0.1-1 M_{\odot}$ -- are excluded at the $>90 $ per cent
C.L. and in some cases violate the total lensing-derived mass limit.
We conclude that this very massive early-type galaxy is dark-matter
dominated inside one effective radius, consistent with the trend recently
found from massive SLACS (Sloan Lens ACS) galaxies, with a total density slope
shallower than isothermal and a IMF normalization consistent with Salpeter.

\end{abstract}

\begin{keywords}
dark matter - galaxies: ellipticals - gravitational lensing:strong -
galaxies: kinematics and dynamics - galaxies: structure - galaxies:
formation
\end{keywords}

\section{Introduction}

Understanding the evolution and the internal structure of massive
early-type galaxies (ETG), as well as their stellar and dark matter
distributions, is crucial in order to fully comprehend the processes
in hierarchical galaxy formation (e.g.\ White \& Rees 1978; Davis et
al. 1985; Frenk et al. 1985). In this context, the relationship
between baryonic matter, which dominates astrophysical observables, and
dark matter (DM), which dominates most of the dynamics during galaxy
formation is particularly important.

To unravel some of the these issues and paint a more robust physical
picture, enormous effort has been afforded to study the relative
contributions of baryonic, dark matter and black hole constituents of
ETGs through stellar dynamical tracers, X-ray studies, and
gravitational lensing (e.g. Fabbiano 1989; Mould et al. 1990; Saglia,
Bertin \& Stiavelli 1992; Bertin et al. 1994; Franx, van Gorkom \&
de Zeeuw 1994; Carollo et al. 1995; Arnaboldi et al. 1996; Rix et
al. 1997; Matsushita et al. 1998; Loewenstein \& White 1999; Gerhard
et al. 2001; Seljak 2002; Borriello, Salucci \& Danese 2003;
Romanowsky et al. 2003; Treu \& Koopmans 2004; Cappellari \& Emsellem 2004; Cappellari et al. 2006; Thomas et al. 2007; Czoske et al. 2008: Czoske, Barnab\'{e} \&
2008; Auger et al. 2010; Treu 2010).

The picture emerging over the last decades from studies of the inner
regions of early-type galaxies, where baryonic and dark matter are
both present, is that to first order the total mass density profile
inside few effective radii can be well-described by a
power-law form close to an isothermal profile with $\gamma
\approx 2$ for $\rho_{\rm tot}=r^{-\gamma'}$, although with a $\sim10$
per cent intrinsic scatter in the density profile (e.g.\ Gerhardt et
al. 2001; Treu \& Koopmans 2004; Koopmans et al. 2006, 2009; Auger et al. 2010, Barnabe
et al. 2008, 2011). The dark-matter density profile in the same
region, however, is far less well constrained although seems
consistent with a density slope $\gamma_{\rm DM} \approx 1.3$
(e.g. Treu \& Koopmans 2004).

In addition, while the innermost regions of early-type galaxies are
expected to be dominated by the stellar mass component, the
dark-matter mass component is usually found to play a non-negligible
role, with mass fractions ranging from 10 to 40 per cent of the total
mass within one effective radius (e.g.\ Gerhard et al.\ 2001;
Cappellari et al.\ 2006; Gavazzi et al.\ 2007; Weijmans et
al.\ 2008). Even more recently, observations, as well as theoretical
studies based on stellar population and dynamical models (e.g. Bullock
et al.\ 2001, Padmanabhan et al.\ 2004), indicate that, for a constant IMF, the dark matter
fraction in the internal region increases monotonically with the mass
of the galaxy (e.g.\ Zaritsky et al. 2006; Auger et al. 2010), a trend
that is more conspicuous in the case of slow-rotator ellipticals
(Tortora et al. 2009).
On the other hand, it has also been shown that the luminous stellar 
mass-to-light ratio scales with the luminous mass of the system (Grillo et al 2010).
However, information from stellar kinematics is limited to the central
regions (a few effective radii) because of the lack of bright
kinematic tracers at large radii, and the total mass determination
suffers from the well-known degeneracy between the mass density
profile of the galaxy and the anisotropy of its stellar velocity
dispersion tensor (Binney \& Mamon 1982). Higher-order velocity
moments, which potentially allow one to disentangle this degeneracy by
providing additional constraints (Gerhard 1993; van der Marel \& Franx
1993), can only be measured with sufficient accuracy in the inner
parts of nearby galaxies with current instruments. Despite great
progress, it still remains difficult to separate the stellar and
dark-matter components, mostly due to a still relatively poor
understanding of the precise shape of the stellar IMF and its
associated stellar mass-to-light ratio. Uncertainties related to the
latter can easily lead to a factor of a few uncertainty in the
inferred stellar mass.
	
Understanding the stellar IMF in massive early-type galaxies is a key open issue with 
a broad range of astrophysical implications. Although it is commonly assumed that the IMF 
is universal and independent of cosmic time, several authors have suggested that the 
IMF might indeed evolve (Dave 2008; van Dokkum 2008) or depend on 
the the stellar mass of the system (e.g. Worthey et al. 1992; Trager et al. 2000; Treu et al. 2010; Graves \& Faber 2010, Auger et al. 2010).
Recently, van Dokkum \& Conroy (2010) have suggested that low-mass
stars ($\la 0.3\,M_{\odot}$) could be far more dominant in massive
early-type galaxies than previously thought. This could imply that
much of the increase in the mass-to-light ratio of galaxies with
galaxy mass is due to a changing stellar IMF rather than an increasing
dark-matter fraction, consistent with the suggestions of Treu et al. 2010 and Auger et al. 2010.

Valuable additional information on distant early-type galaxies,
besides their kinematics, can be provided by gravitational lensing.
Strong gravitational lensing provides a very useful tool to
investigate the total mass content of the lens (Maoz \& Rix 1993;
Rusin \& Ma 2001; Rusin \& Kochanek 2005; Dye \& Warren 2005; Brewer
\& Lewis 2006b, 2008) and to place constraints on the inner mass
distribution of lens galaxies at redshift out to $z \simeq 1$.
Unfortunately, the mass-sheet and the related mass-profile (Wucknitz
2002) degeneracies do not always allow one to accurately determine the
slope of the galaxy density profile and above all to unambiguously
disentangle the luminous and dark matter components.

Gravitational lensing and stellar dynamics are particularly effective
to break many of these degeneracies, when they are applied in
combination in the analysis of the internal structure of distant
early-type galaxies.  The two approaches are complementary and allow,
when combined self-consistently, to robustly determine the mass
profile of the lens galaxy (Koopmans \& Treu 2002, 2003; Treu \&
Koopmans 2002, 2003, 2004; Jiang \& Kochanek 2007; Barnab\'{e} \&
Koopmans 2007; Czoske et al. 2008; Czoske, Barnab\'{e} \& 2008;
Koopmans et al. 2009; Grillo et al. 2010; Treu 2010; Barnab\'{e} et
al. 2011). Disentangling stellar and dark matter remains difficult,
but the uncertainties have been reduced from factors of a few to far
less than this (Treu et al.\ 2010).

\ctable[
caption = Properties of the cosmic Horseshoe$^{1}$.
]{llr}{
\tnote[1]{Belokurov et al.\ (2007) measured the redshift of the source to be $z=2.379$. We find a systematic shift that brings the source redshift to be $z=2.3811$ in agreement with Quider et al. (2009).}  
\tnote[2]{The mass within the Einstein radius or, more precisely, within the ring diameter, is taken from Dye et al.~(2008). }
\tnote[3]{Parameters obtained from images taken with the 2.5 m Isaac Newton Telescope (INT). Magnitudes are taken from SDSS DR7. See Belokurov et al. (2007)}
}{
\hline
& \begin{small} \textbf{Parameter} \end{small} & \begin{small} \textbf{Values} \end{small}\\
\hline
\begin{small} \textbf{Lens} \end{small} &\begin{small} RA \end{small}& \begin{small}11h 48m 33.15s\end{small} \\
\begin{small} \textbf{Galaxy} \end{small} &\begin{small} Dec \end{small} & \begin{small}19$^{\circ}$ 30\arcmin\ 03\farcs5\end{small} \\
& \begin{small} Redshift \end{small}& \begin{small}0.444\end{small} \\
&  \begin{small} Effective radii  \end{small}& \begin{small}$1\farcs 96 \pm 0\farcs 02 $\end{small}   \\
& \begin{small} $g_{L}$ \end{small}& \begin{small}$(20.84\pm0.06)$~mag\end{small} \\
& \begin{small} $r_{L}$ \end{small}& \begin{small}$(19.00\pm0.02)$~mag\end{small} \\
& \begin{small} $i_{L}$ \end{small}& \begin{small}$(18.22\pm0.01)$~mag\end{small} \\
& \begin{small} $z_{L}$ \end{small}& \begin{small}$(17.75\pm0.04)$~mag\end{small} \\
& \begin{small} Axis ratio, $g$  \end{small}&\begin{small} $0.8\pm 0.1$ \end{small}\\
\hline
\begin{small} \textbf{Source} \end{small} & \begin{small} Redshift$^{2}$ \end{small} & \begin{small}$2.38115 \pm 0.00012$\end{small} \\
&\begin{small}  Star formation rate\end{small}& \begin{small}$\sim100\,{M}_{\odot}\,{\rm yr}^{-1}$\end{small}\\
& \begin{small} Dynamical mass \end{small}& \begin{small}${M}_{vir}\simeq10^{10}~{M}_{\odot}$\end{small}\\
\hline
\begin{small} \textbf{Ring} \end{small} & \begin{small} Diameter \end{small}& \begin{small} 10\farcs2 \end{small}\\
& \begin{small} Length \end{small}& \begin{small} $\sim300^{\circ}$\end{small} \\
&  \begin{small} $u_{L}$\end{small}& \begin{small}21.6 mag \end{small}\\
&  \begin{small} $g_{L}$ \end{small}& \begin{small}20.1 mag \end{small}\\
&  \begin{small} $i_{L}$ \end{small}& \begin{small}19.7 mag \end{small}\\
&  \begin{small} Mass enclosed$^{3}$ \end{small}& \begin{small} $(5.02 \pm0.09)\times 10^{12}\,M_{\odot}$\end{small}\\
\hline
}

\begin{figure} 
\center
\includegraphics[height=8cm]{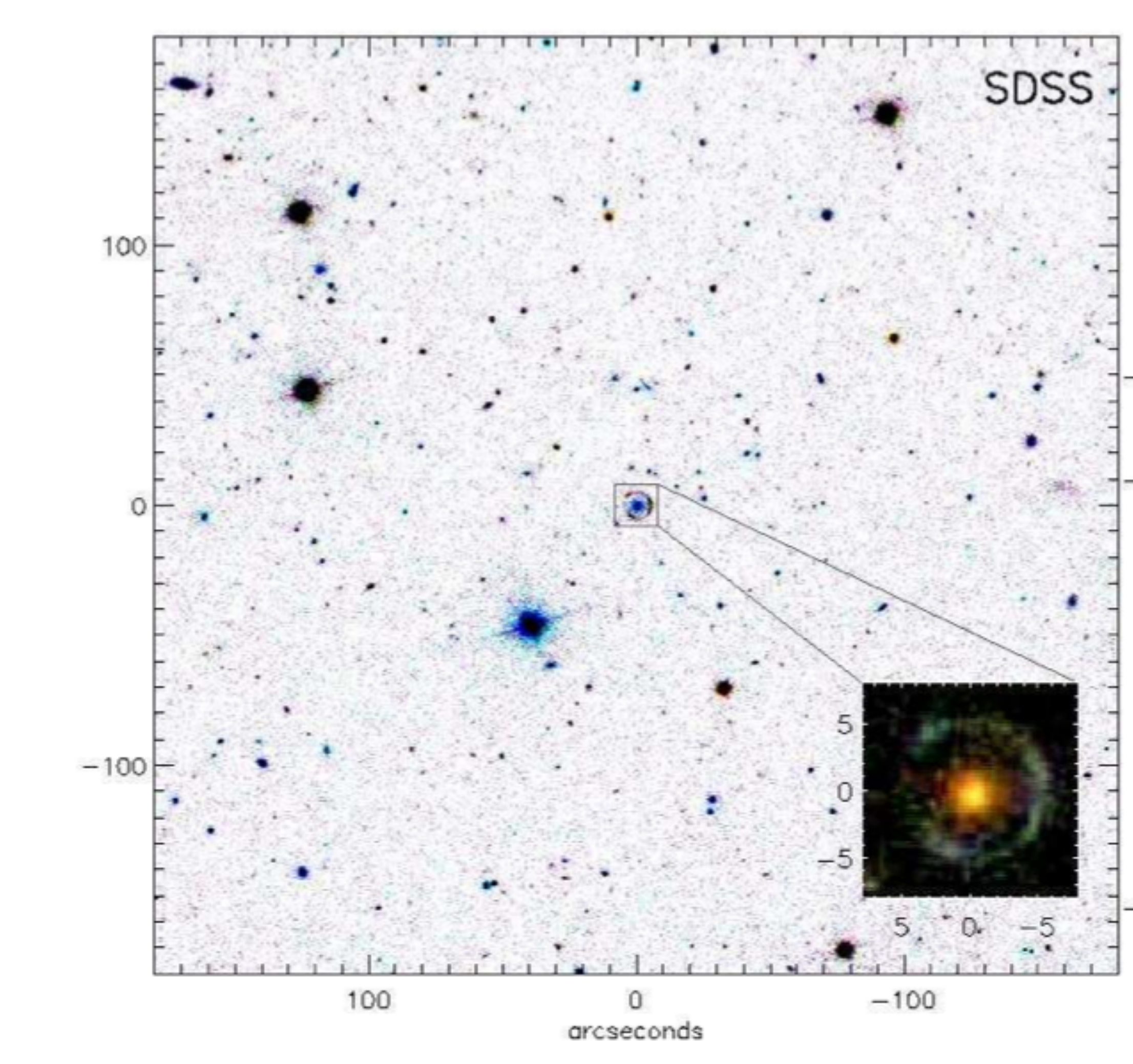}
\caption{SDSS color composite image of  the sky region around the Cosmic Horseshoe (the zoom is $16\arcsec\times 16\arcsec$).}
\label{fig:sdss}
\end{figure}

The X-Shooter Lens Survey (XLENS) aims to take the next step and spectroscopically observe
 a sample of lens galaxies from the Sloan Lens ACS Survey (SLACS, Bolton et al. 2006) 
 with $\sigma_{\rm ETG}\geq 250\,\kms$ using the X-Shooter instrument\footnote{X-Shooter is a
  powerful broad-band (3000--25000~ \AA), three-armed medium-resolution
  spectrograph on the VLT (D'Odorico et al.\ 2006); $22^{h}$ GTO time
  on three systems (084.A-0289 and 087.A-0620) and $40^{h}$ GO time
  were awarded (P086.A-0312) to the XLENS project (PI: Koopmans)}.
With this data we intend to further disentangle the stellar and
dark-matter content of the galaxies, through combined lensing,
dynamical and {\sl spectroscopic} stellar population studies.  With
multi-band HST images in hand, we are able to obtain more precise
dark-matter mass fractions than ever before, in order to ultimately
correlate these with ETG mass and compare with theoretical galaxy
formation models.
By combining the lensing and dynamical results with stellar population 
models we will plan to constrain the normalization and shape of the stellar initial mass function.

In this paper we present the result of a pilot program of the XLENS
project: A study of the ``Comic Horseshoe'' (SDSSJ1148+1930), an
almost complete Einstein ring with a diameter of $\sim10\farcs 2$
around a very massive early-type galaxy at $z=0.444$. The system was
discovered by Belokurov et al.\ (2007) in the Sloan Digital Sky Survey
Data Release 5 (DR5, Adelman-McCarthy et al., 2007) . 
The source is a star-forming galaxy at $z=2.381$
(Dye et al. 2008; Quider et al. 2009).  Properties and characteristics
of the Cosmic Horseshoe are listed in Table~1 and a composite SDSS
image of the system is shown in Figure~\ref{fig:sdss}.

The paper is organized as follows: In Section 2, we present the
observations and data reduction. In Section 3, we discuss our
kinematic analysis. In Section 4, we discuss the luminous and dark
matter distributions of the lens galaxy. We summarize our findings and
we present our conclusions in Section 5.  We assume $H_{\rm 0}=70
\,\mathrm{km \, s^{-1}\,Mpc^{-1}}$, $\Omega_{\rm m}=0.3$ and
$\Omega_{\Lambda}=0.7$ throughout the paper.

\section{OBSERVATIONS AND DATA REDUCTION}

X-Shooter observations of SDSSJ1148+1930 were carried out during a GTO
run between 17--24 March 2010 in slit mode\footnote{P084.A-0289(A);
  PI: Koopmans}, splitting the beam over three arms: UVB (R=3300 with
$1\farcs6$ slit); VIS (R=5400, with $1\farcs5$ slit); and NIR (R=3300
with $1\farcs5$ slit), covering a wavelength range from 3000 to
25000 \AA\ simultaneously.  The 11\arcsec\ long slit was centered on
the galaxy with a position angle (PA) of 163$^{\circ}$. The latter
minimizes contamination from the source and leaves enough sky region
to facilitate accurate sky subtraction.  Two Observation Blocks (OBs) were not used because
of bad seeing and/or an incorrect positioning of the slit on the
object.  The total exposure time on target for each arm is $\sim 7389$
sec and the typical seeing is $\sim0\farcs6$.  Standard calibration
frames were obtained during daytime after the corresponding OB.  A
summary of the observing blocks is given in Table 2.
 
\begin{figure*}  
\includegraphics[height=8.2cm]{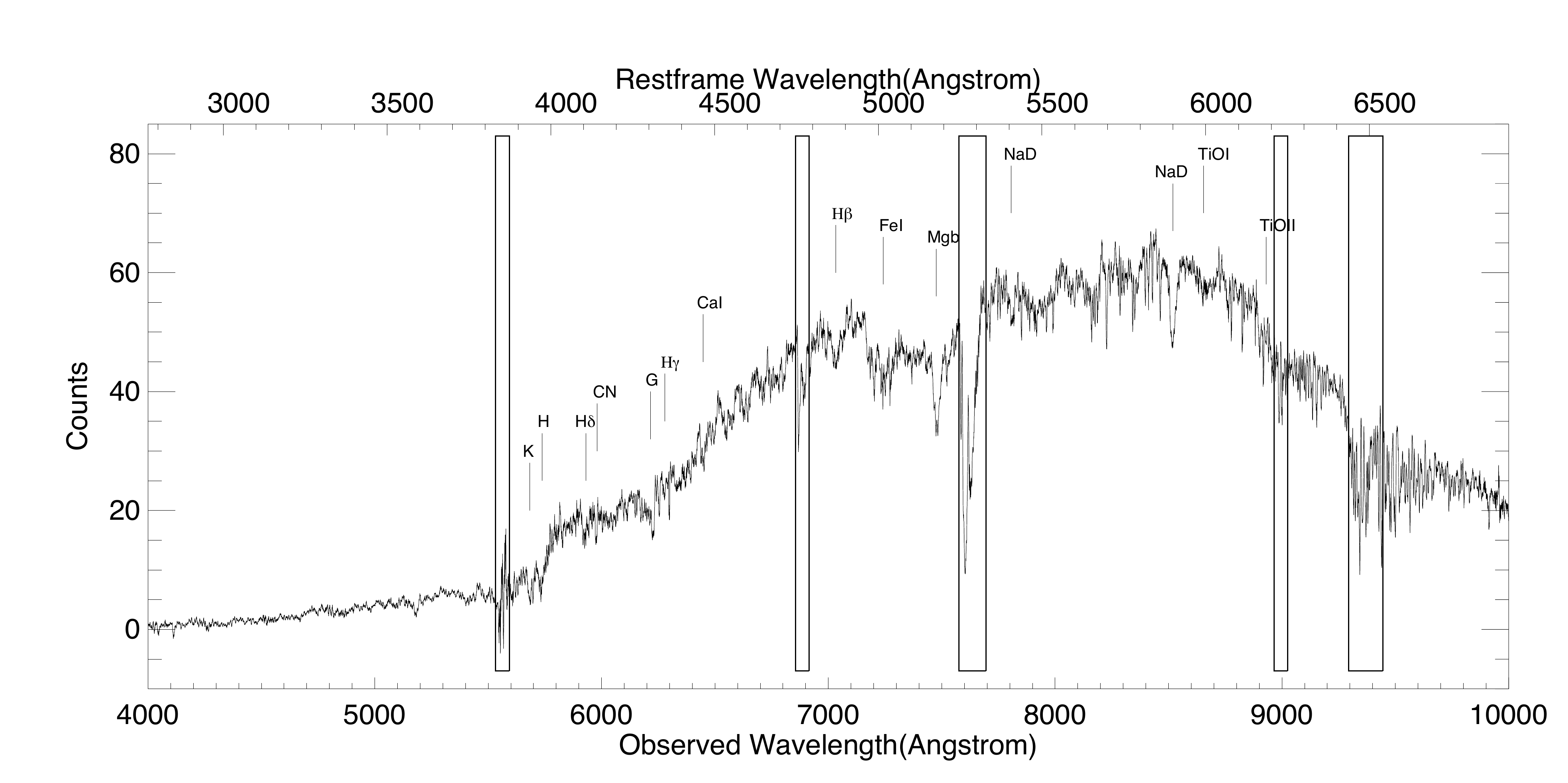}
\caption{The final luminosity-weighted UVB--VIS 1D X-Shooter spectrum
  extracted from a rectangular aperture of $1\farcs8 \times 1\farcs6$
  centered on the galaxy (see text). Spectral features the lens galaxy
  are marked.  Telluric absorption lines have not been removed from
  the spectrum nor has the spectrum been flux calibrated. 
  The boxes indicate parts of spectrum that are affected
  by sky or telluric lines.}
\label{fig:finspec}
\end{figure*}

Data reduction was done using the ESO X-Shooter pipeline v1.2.1
(Goldoni et al. 2006) and the Gasgano data file organiser developed
by ESO.  The pipeline reduction uses calibration spectra, taken during
the commissioning run, for bias subtraction and flat-fielding of the
raw spectra. Cosmic rays are removed using LACosmic (van Dokkum 2001).
For each arm, the orders are extracted and rectified in wavelength
space using a wavelength solution previously obtained from the
calibration frames.  The resulting rectified orders are shifted and
coadded to obtain the final two-dimensional (2D) spectrum.  We extract
a one dimensional spectrum (1D) from the resulting 2D merged spectrum,
using an IDL code that uses the optimal-extraction algorithm
of Horne (1986).  It also produces the corresponding error file and
bad pixel map.  The final signal-to-noise ratio in the UVB+VIS
spectrum is $\sim 7 $ per pixel.  No telluric correction was applied,
so that prominent atmospheric absorption bands can still be seen in
the final spectrum (Fig \ref{fig:finspec}).

Because the near-infrared spectrum suffers seriously from sky-line
residuals using the current pipeline, we limit our analysis to the
UVB-VIS region of the spectrum and defer a full analysis of the
infrared data as well as our spectroscopic stellar-population analysis
to future work. In this paper, we base our stellar mass determinations
solely on the broad-band colors from the SDSS plus the stellar
kinematic and lensing data.

 \begin{table}
 \center
  \caption{Observational details of the Cosmic Horseshoe}
\label{tab:observation}
 \begin{tabular}{llll} 
\hline
\textbf{Observation } &\textbf{Date} & \textbf{Exp. time} & \textbf{Seeing} \\
\textbf{Block } &\textbf{ } & \textbf{(sec)} & \textbf{ } \\

\hline
 200200337 & 3-17-2010 & 3 x 821(UVB - VIS) &  $0\farcs62$ \\
  & 											 & 3 x 3 x 274(NIR)  &\\
 200200343 & 3-17-2010 & 3 x 821(UVB - VIS) & $0\farcs56$ \\
  & 											 & 3 x 3 x 274(NIR)  &\\
 200200436 & 3-19-2010 & 3 x 821(UVB - VIS) & $0\farcs66$ \\
  & 											 & 3 x 3 x 274(NIR)  &\\
\hline
\end{tabular}
\end{table}

\section{Stellar Kinematics}
\begin{figure*}  
\includegraphics[height=13cm]{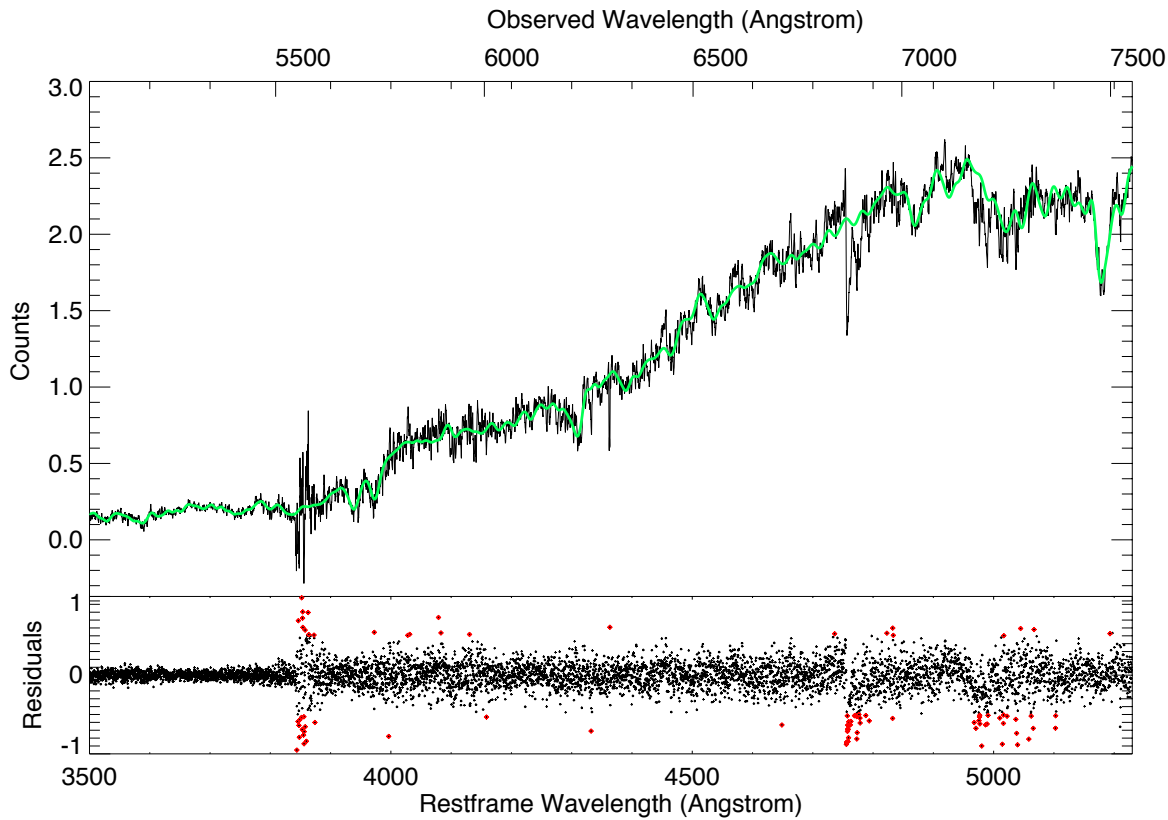}
\caption{\textit{Top panel:} Logarithmically rebinned UVB+VIS galaxy
  spectrum (black) and logarithmically rebinned best-fit template
  (green), both in restframe wavelength. \textit{Bottom panel:}
  Residuals from the fit. Bad-pixels excluded from the fitting
  procedure (sky line, telluric lines) are shown in red. See the text
  for more information.}
\label{fig:ppxf}
\end{figure*}

We measure the luminosity-weighted velocity
dispersion (LOSVD) of the lens galaxy from the final 1D UVB--VIS spectrum
using the Penalized Pixel Fitting (PPxF) code of Cappellari \&
Emsellem (2004).  PPxF determines the combination of stellar templates
which, when convolved with an appropriate line-of-sight-velocity
distribution, best reproduces the galaxy spectrum.  The best-fitting
parameters of the LOSVD are determined by minimizing a $\chi^{2}$
penalty function.  The best fit provides the mean velocity and the
velocity dispersion ($v$ and $\sigma$, respectively), plus their uncertainties. 
The S/N ratio of the data is inadequate to measure
the higher order Gauss-Hermite moments $h_{3}$ and $h_{4}$, which
quantify the asymmetric and symmetric departures of the LOSVD from a
pure Gaussian (related to the skewness and kurtosis respectively).
PPxF also allows the user to mask noisy or bad regions of the galaxy
spectrum.  We also perform an iterative sigma-clipping in order to
clean the spectrum of residual bad pixels, sky lines and cosmic rays.
We focus on absorption lines between 3500--5500 \AA\ (including Ca K
and H, G4300, H$_{\beta}$, Mg$_{b}$ and some Fe lines).  To minimize
errors due to mismatch between the resolution of the templates and the galaxy spectrum, 
we use X-Shooter stellar spectra obtained as part of the
X-Shooter Stellar Library (XSL) survey (Trager et al.\ 2011 in prep.),
with similar instrumental resolution (for the galaxy spectrum we use a $1\farcs5$ slit, corresponding to
 $\sigma_{\rm instr}\sim25\,\kms$, while for the stellar templates the slit width is  $0\farcs7$,corresponding to  
 $\sigma_{\rm instr}\sim12\,\kms$).
An excellent fit is obtained with a weighted linear combination of a
K1 giant template (57 per cent) and a G2 star template (43 per cent).
The selected region of our galaxy spectrum used for the
fit and the corresponding best fit stellar template are shown in
Figure \ref{fig:ppxf}.

\subsection {Luminosity-Weighted Kinematics} 

The measured luminosity-weighted velocity dispersion\footnote{The only
  velocity dispersion value previously published for the Cosmic
  Horseshoe is from Belokurov et al.\ (2007). They perform fits of
  Gaussian line profiles to the Ca H and K absorption lines from a
  much lower resolution spectrum (FWHM$\sim10$ \AA).  They find a
  higher velocity dispersion estimation of $430\pm50\,\kms$,
  inconsistent with our high-quality data.} for the central aperture
of $1\farcs8 \times 1\farcs6$ is $\langle \sigma \rangle = 352 \pm 10
\pm 16 \,\kms$.  The formal error on the dispersion includes both the
random error contribution and the systematic uncertainties due to
spectral range differences, template mismatch and continuum fitting as
discussed below.  Figure \ref{fig:ppxf} shows the best fitting
template of the PPxF routine superimposed on the galaxy spectrum as
well as the residuals of the fit.

As a first test of the accuracy of our measurements, we use the more
heterogeneous
MILES\footnote{http://www.iac.es/proyecto/miles/pages/stellar-libraries/miles-library.php}
stellar template library (S\'{a}nchez-Bl\'{a}zquez, et al., 2006).  We
select 100 stars (F, G, K, M) in the range 3525--7500 \AA, with
2.3 \AA\ FWHM spectral resolution.  The measured luminosity-weighted
stellar velocity dispersion of $\langle \sigma \rangle = 358 \pm
31\,\mathrm{km\,s^{-1}}$, after instrumental correction, is consistent
with the above estimate, based on XSL templates, but has larger errors
due to the lower resolution of the MILES library. As a second test, we
fit templates to two different spectral regions. We find slightly
different results between the blue and the red part of the spectrum,
but always consistent within $2\sigma$ (not including systematics). 
The scatter in these fits is used to estimate additional systematic uncertainties related to
template mismatches and spectra coverage.

\subsection{Spatially-Resolved Kinematics}

\begin{figure*}  
\includegraphics[height=6.5cm]{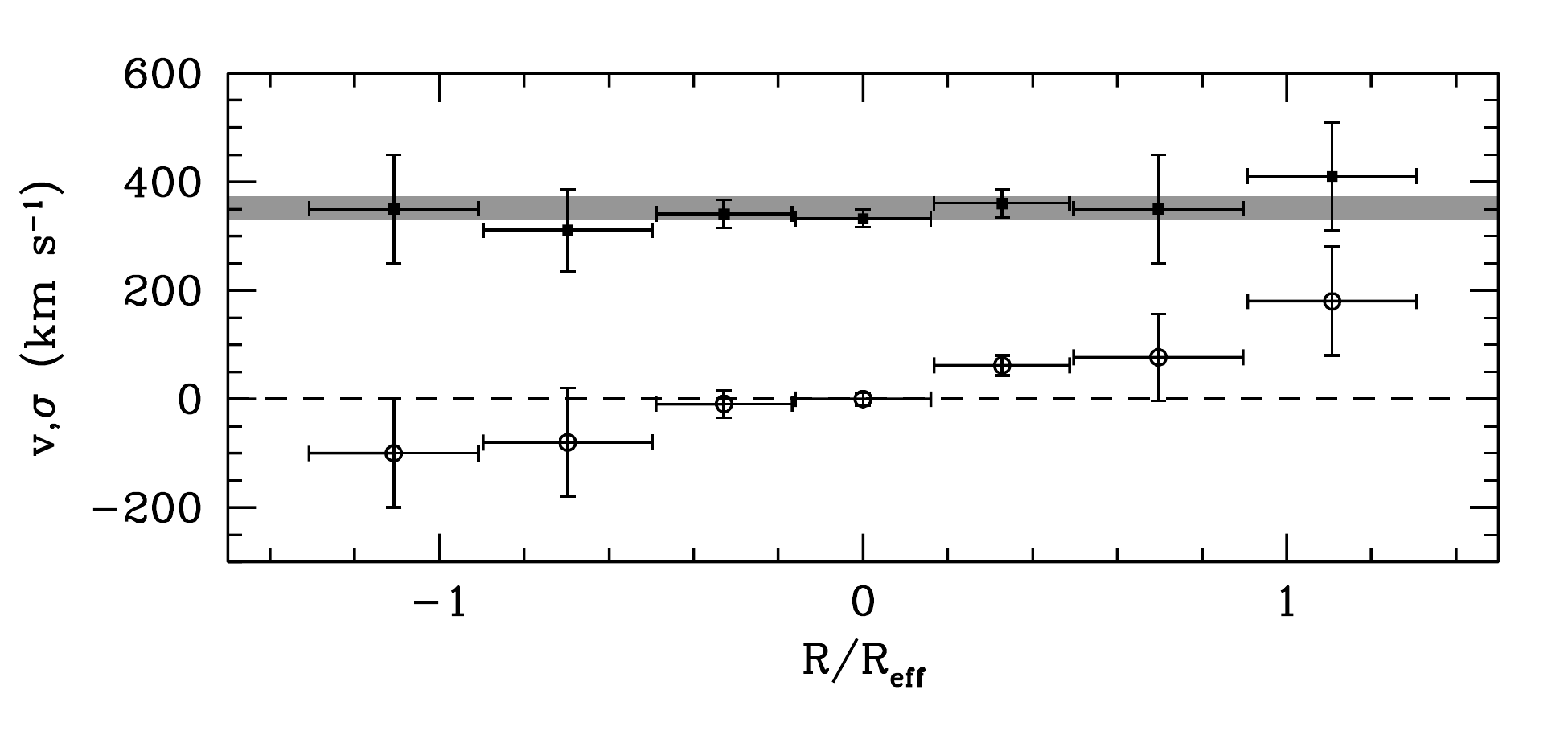}
\caption{Kinematic profiles of the Cosmic Horseshoe. Rotation (bottom)
  and velocity dispersion (top). Grey line shows the light-weighted
  value for the central aperture.}
\label{fig:spatially}
\end{figure*}

To preserve the spatially-resolved kinematic information, we define
seven spatially-varying apertures (with adequate S/N ratio) along the
radial direction and we sum the signal within each aperture.
Apertures are defined to be larger than the seeing, in order to have
independent kinematics measurements for each aperture.  The stellar
rotation velocity and velocity dispersion are measured in each
aperture using PPxF, as described above. Again, we use different
spectral regions, excluding the most prominent telluric lines in the
VIS range and a range of seven XSL stellar templates (G, K and M
stars).  The uncertainties on the inferred kinematics are estimated by
adding in quadrature the formal uncertainty given by PPxF and the
scatter in the results for different templates and spectral regions.
Details of the aperture sizes and the kinematic profiles are listed in
Table 3.  The rotation and the velocity dispersion profiles are shown
is Figure \ref{fig:spatially}.

We find an almost flat velocity dispersion profile beyond the
effective radius.  The weighted average value of $344 \pm 25\,\kms$ is
consistent within the formal error with the luminosity weighted value
for an aperture of $1.8\arcsec \times 1.6\arcsec $ (see Table 3), as
expected.  The velocity profile shows some mild rotation, with a
projected rotation velocity of $\sim 140\,\kms$ at one effective
radius.  Because the effective dispersion ($v^{2}_{\rm
  rms}=\sqrt{v^{2} + \sigma^{2}}$; see Cappellari 2008) is well within
the errors on $\sigma$, the effect of rotation can be neglected in our
spherical Jeans analysis and we will ignore rotation in the remainder
of this paper.

\section{Stellar and Dark-Matter}

Here we derive the slope of the total density profile and calculate
the fraction of dark matter inside one effective radius.
 
\subsection{The Galaxy Mass Model}

To derive the stellar mass inside the Einstein radius, we follow Treu
\& Koopmans (2002), Koopmans \& Treu (2003), Treu \& Koopmans (2004).
We model the mass distribution as a superposition of the stellar and
the dark matter components.  For the stellar mass distribution we use
two different spherical models, described by the equation:
\begin{equation}
\rho_{\rm *}(r)= \dfrac{ (3-n)M_{*} r_{*}}{4\pi r^{n} (r + r_{*})^{4-n}}
\end{equation}
The Hernquist (1990) luminosity-density profile has $n=1$,
and the Jaffe (1983) luminosity-density profile has $n=2$. The 
 total stellar mass is $M_{*}$ and the break-radius is $r_{*}$. 

We model the dark matter distribution as a generalized spherical NFW
profile, using a break radius of $r_{b}=50$\,kpc, is typical of
massive ETGs (although the result is insensitive to its exact value)
and an outer slope of $\gamma_{\rm out}=3$ (Gavazzi et al. 2007; Auger et al. 2010b), 
following the prediction from numerical simulations of dark matter halos (Ghigna et al., 2000):
\begin{equation}
\rho_{\rm DM}(r)= \dfrac{ \rho_{d,0}}{ (r/r_{b})^{-\gamma} [1+(r /r_{b})^{2}]^{(\gamma-3)/2}}
\end{equation}
The break radius and the density scale ($\rho_{d,0}$) determine the
virial mass of the dark matter halo (Bullock et al., 2001).

For completeness, we also use a single mass component model where the
stellar and dark-matter mass densities add to a power-law with an
effective density slope $\gamma'$ for $\rho_{\rm tot} \propto
r^{-\gamma'}$ and the stars are treated as trace particles (see
Koopmans \& Treu 2003).

The most accurately known constraint from the lens mass model is the
mass inside the Einstein radius.  We use $M_{\rm E} = (5.02 \pm 0.09)
\times 10^{12}\,M_{\odot}$ inside $\theta_{\rm E}= 5\farcs1$ derived
by Dye et al. (2008). The mass components must add to
$M_{\rm E}$ within the Einstein radius.  The error in radius is folded
into the error in mass so that a fixed radius can be used (the mass
and Einstein radius are coupled in the modeling).  We use the average
of the two effective radii from the $g$ and $i$ band images reported
in Belokurov et al.\ (2007), $R_{\rm eff}=1.96 \pm 0.02$\,arcsec
(Table 1). The effective radius is uniquely related to $r_{*}$ (see
Hernquist 1990 and Jaffe 1983).

We also assume a constant orbital anisotropy parameter $\beta$ and
allow it to range between 0.0 and 0.5, which is typical for massive
early-type galaxies (e.g. Gerhard et al.\ 2001).  The parameters of
the model without a strong prior are then the stellar mass ($M_{*}$)
and the dark-matter density slope ($\gamma$), or only $\gamma'$ for
the single-component model.

We subsequently solve the spherical Jean equations and compare the
models to the kinematic data, taking the aperture sizes and seeing
into account. We find an effective density slope of $$\gamma' =
1.72_{-0.06}^{+0.05} {\rm ~~(68 per cent~C.L.)}$$ when marginalizing
over $\beta=0.0$--$0.5$, shallower than isothermal (see Koopmans et
al.\ 2006 and 2009).  
This slightly low value for the logarithmic total-density slope may suggest 
that this object can be a group, or a small cluster of galaxies, 
where the overall efficiency of converting gas into stars is lower, 
and where typically the overall mass density profile in the corresponding 
region is shallower than isothermal (e.g. Newman et al. 2011).
Changing the luminosity-density profile also
does not change the final logarithmic slope significantly.  The more
interesting case of the two component model will now be discussed.

\begin{table}
 \label{tab:spatially}
  \caption{Spatially resolved kinematics of the Cosmic Horseshoe.}
  \begin{tabular}{rrcc} 
\hline
\textbf{Aperture}  &\textbf{Aperture} & \textbf{$v$} &\textbf{$\sigma$} \\
\textbf{center (\arcsec)}  &\textbf{dimension(\arcsec)} &\textbf{($\mathrm{km\,s^{-1}}$)} & \textbf{($\mathrm{km\,s^{-1}}$)} \\
\hline
0.00, 0.00 & $1.80 \times1.60$ & 0 $\pm$ 15 & 352 $\pm$ 10 \\ 
\hline
$-2.16, 0.00$ & $0.80 \times1.60$  & $-100$ $\pm$ 100 & 350 $\pm$ 100 \\ 
$-1.36, 0.00$ & $0.80 \times1.60$ & $-80$ $\pm$ 100 & 311 $\pm$ 76 \\ 
$-0.64, 0.00$ & $0.64 \times1.60$  & $-9$ $\pm$ 25 & 341 $\pm$ 26 \\ 
$0.00, 0.00$ & $0.64 \times1.60$ & 0 $\pm$ 12 & 332 $\pm$ 16 \\ 
$+0.64, 0.00$ & $0.64 \times1.60$  & 62 $\pm$ 18 & 360 $\pm$ 25 \\ 
$+1.36, 0.00$ & $0.80\times1.60$  & 77 $\pm$ 80 & 350 $\pm$ 100 \\ 
$+2.16, 0.00$ & $0.80\times1.60$  & 180 $\pm$ 100 & 410 $\pm$ 100 \\ 
\hline
\end{tabular}
\end{table}

\subsection{The Stellar Mass Fraction from Lensing and Kinematic Constraints}

To derive the stellar mass fraction inside the Einstein radius, we
create a densely sampled grid of likelihood values by comparing the
kinematic profiles to the data for a projected stellar mass fraction
[$f^{*}(<R_{\rm E})$] within the Einstein radius ranging between 0 and
1 and a dark-matter density slope ($\gamma$) ranging between 0.0 and
2.0. We assume flat priors on both quantities and marginalize over
$\gamma$ to derive the probability density function of $f^{*}(<R_{\rm
  E})$. The results for the Hernquist and Jaffe profiles are shown in
Figure~5 for $\beta=0$ and a more extreme case of radial anisotropy
with $\beta=0.5$.
Taking as reference the best fit parameters of the Sersic profile computed in Dye et al.(2008): $n=5.40 \pm 0.04$, $r_{0}=3.9 \pm 0.1$, $L_{1/2}=61.2\pm 0.4$), 
with the form:
\begin{equation}
L = L_{1/2} exp \lbrace -B(n) [(r/r_{0})^{1-n} -1] \rbrace 
\end{equation} 
we conclude that Hernquist and Jaffe profiles fit well the observed luminosity profile 
of the galaxy within the effective radius (with a slight preference for the Hernquist model).
\\
We find a fraction of stellar mass within the Einstein radius for the
two luminosity profiles of:
\begin{equation}
f^{*}_{\rm HQ}= 0.19 ^{+0.04}_{-0.09} {~~\rm and ~~}
f^{*}_{\rm JF}=0.13 ^{+0.03}_{-0.05} {~~\rm for ~~} \beta = 0 \nonumber
\end{equation} 
and 
\begin{equation}
f^{*}_{\rm HQ}= 0.13 ^{+0.04}_{-0.07} {~~\rm and ~~}
f^{*}_{\rm JF}=0.11 ^{+0.02}_{-0.05} {~~\rm for ~~} \beta = 0.5 \nonumber
\end{equation} 
We note that in projection, the lens galaxy is already fully
dark-matter dominated inside $\sim 2.5$ effective radii.  For
comparison with previous work, we also derive the dark-matter fraction
inside the effective radius and find  $$f_{\rm DM}(<R_{\rm
  eff}) = 0.60^{+0.15}_{-0.06} \pm 0.1,$$ [68 per cent confidence
  level (CL)] for $\beta=0$, including a systematic uncertainty of 0.1. The random error
is based on the marginalized probability densities shown in Figure~5
and a systematic error is included based on the maximum range of
possible dark-matter density slopes. Although the latter is
constrained by the models, we extrapolate inward from $R_{\rm E}$ to
$R_{\rm eff}$, where this slope could be somewhat different. We note
though that the systematic error is rather conservative. The
dark-matter fraction increases by $\sim 0.1$ for $\beta=0.5$ and the
difference between the Hernquist and Jaffe profiles is negligible (by
construction, since they both contain equal fractions of mass inside
that radius). This high dark-matter fraction inside the effective
radius is consistent with the result found in Auger et al. (2010) for
SLACS systems and is consistent with the dark matter fraction 
within the effective radii beeing a monotonically-increasing function of galaxy mass.

\begin{figure*}
\includegraphics[height=12.7cm]{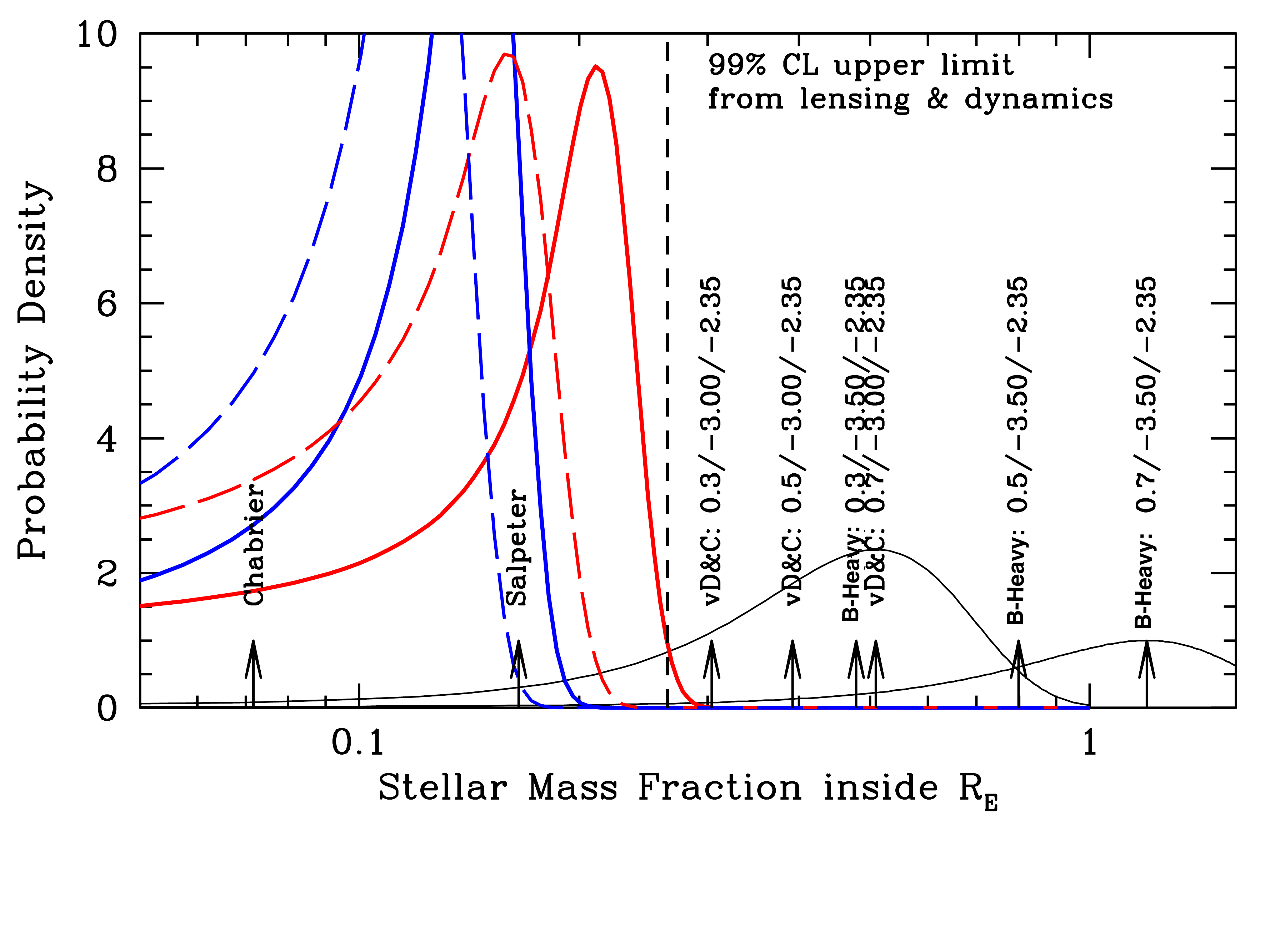}
\caption{Marginalised PDF for the Hernquist model (red lines) and the
  Jaffe model (blue lines) from our lensing and dynamics model. Solid lines are for an anisotropy parameter of  $\beta=0.0$, 
  and dashed lines for $\beta=0.5$. Black arrows show the internal stellar
  mass fractions with different IMFs from our stellar population analysis. Black lines represent PDFs of Van Dokkum \& Conroy IMF and of the most extreme  bottom-heavy IMFs. In the labels, the first number represents the break radius, the second is the IMF slope at lower masses ($m \le m_{break}$) and the last number is the slope at higher masses (fix to Salpeter slope).
  We exclude the Chabrier IMF and the $\alpha=-3$ IMFs at a
  confidence level of $\geq90$ percent and the extreme
  bottom-heavy ($\alpha=-3.5$) IMFs at a
  confidence level of 99 percent, while the Salpeter IMF is fully
  consistent with lensing data.}
\label{fig:imf}
\end{figure*}

\subsection{Stellar Mass Fraction from Stellar Population Constraints}

We independently calculate the projected stellar mass fraction inside
the Einstein radius using stellar population synthesis models and the
SDSS colors (see Table~1) of the lens galaxy. A comparison between
this fraction and that derived from lensing and stellar kinematics
provides a direct constraint on the stellar IMF (see also e.g.\ Treu
et al. 2010; Auger et al. 2010).

We use GALAXEV, a library of evolutionary stellar population synthesis
models computed using the isochrone synthesis code of Bruzual \&
Charlot (2003).  This code allows one to compute the spectral
evolution of stellar populations for a wide range of ages and
metallicities.  Here we use the 2007 version of GALEXEV kindly
provided by S. Charlot (a version commonly known as `CB07').  We use
six Simple Stellar Populations (SSP) models computed with a Salpeter
IMF or a Chabrier IMF for a range of metallicities from $Z=0.0001$ to
$Z=0.05$, all of them with lower mass cutoff of $M_{\rm
  low}=0.1$\,M$_{\odot} $ and upper mass cutoff of $M_{\rm
  up}=100$\,M$_{\odot} $.  The spectral resolution is 3 \AA \,across
the wavelength range of 3200--9500 \AA.

We compute the spectral evolution of the stellar population and the
redshift dependence of colors and magnitudes in the SDSS filters $g$,
$r$, $i$, and $z$ for all SSP models and for a range of ages and star
formation histories (SFH).  For the SFH model, we use an exponentially
declining star formation rate (SFR) with time-scales $\tau$ = 0.1 to
0.4 Gyr and as extreme cases an instantaneous burst with $\tau = 0$ or
a constant SFR.  We compute the redshift evolution of the galaxy
colors and magnitudes for a start of star formation between 12 to 5
Gyr, corresponding to a formation redshift of between roughly $z\sim5$
and $\sim0.5$.  By comparing the model magnitudes to the SDSS
magnitudes, we subsequently determine a grid (as function of age,
metallicity and SFH duration) of likelihood values for each model as
well as the total stellar mass.  We do this for both the Salpeter and
Chabrier IMFs.

\begin{table*}
 	\caption{The stellar mass fractions within the Einstein Radius
          derived from lensing plus stellar kinematics and from
          stellar population synthesis models, respectively. The
          likelihood ratio compares maximum a-posteriori difference
          between the SSP model and the no-difference
          (null-hypothesis) model with the isotropic Hernquist model,
          which has the highest inferred stellar mass fraction. The
          IMF slope is assumed to be $\alpha=-2.35$ beyond the break
          $m_{\rm break}$ in the mass function. The probabilities in
          between parentheses represent for each IMF the fraction probability that a given PDF ($P_{0}$)
          matches the max probability.}
 	\label{tab:stmass}		
\begin{tabular}{lccc}
\hline
\textbf{Lensing and Kinematic Model}  & \textbf{Anisotropy}  & {\bf Stellar Mass Fraction} & \\
\hline\\
Hernquist &  $\beta = 0.0$ & $0.19^{+0.04}_{-0.09}$ & \\ 
\\
 & $\beta = 0.5$  & $0.13^{+0.04}_{-0.07}$  & \\ 
\\
Jaffe & $\beta = 0.0$ & $0.13^{+0.03}_{-0.05}$ & \\ 
\\
& $\beta = 0.5$ & $0.11^{+0.02}_{-0.05}$ & \\ ~\\
\hline
&$\mathbf{m_{\rm break}}$  & &$\mathbf{ 2 \ln(P_{\rm max}/P_{0})}$\\
\textbf{Stellar Population Model} & $(M_{\odot})$ & {\bf Stellar Mass Fraction} & (HQ versus SSP)\\
\hline\\
Chabrier & -- &  $0.07 \pm 0.02$ & $3.0~(0.08)$\\
Salpeter  & -- &  $0.17 \pm 0.06$ & $0.1~(0.75) $\\ 
van Dokkum \& Conroy  $\alpha=-3.0$  & 0.3 &$0.30 \pm 0.11$  & $ 1.0~(0.32)$ \\
van Dokkum \& Conroy  $\alpha=-3.0$  & 0.5 & $0.39 \pm 0.15$ & $ 1.9~(0.17)$\\
van Dokkum \& Conroy  $\alpha=-3.0$  & 0.7 & $0.51 \pm 0.18$ & $ 3.0~(0.08) $\\
Bottom-heavy $\alpha=-3.5$ & 0.3 & $0.48 \pm 0.17$ & $ 2.8~(0.09)$\\
Bottom-heavy $\alpha=-3.5$ & 0.5 & $0.80 \pm 0.29$ & $ 4.5~(0.03)$\\
Bottom-heavy $\alpha=-3.5$ & 0.7 & $1.19 \pm 0.43$ & $ 5.0~(0.03)$\\ ~ \\
\hline
\end{tabular}
\end{table*}

We use the standard Bayesian approach, as outlined in Auger et
al.\ (2009), to determine the posterior probability distribution
function and the marginalized errors on the total stellar mass of the
galaxy, assuming flat priors on all parameters in logarithmic space
(e.g. a prior $ \propto 1/\tau$ for $\tau$). 
The latter assumption is
not critical, but given that the time-scale of the SFH, etc., are
unknown \emph{a priori}, this prior is a better description of our
ignorance before making the observations (and modeling).  From the
resulting cumulative probability function, we calculate the median of
the total stellar mass and its 68 per cent confidence interval for
both IMFs. Using the observed brightness profile of the galaxy modeled
as a Hernquist, Jaffe or deVaucoulers profile (the precise choice is
not critical), we determine the fraction of the stellar mass (assuming
a constant $M/L$ ratio) within the Einstein radius. The results of
this analysis are listed in Table \ref{tab:stmass}, where we report
the inferred total stellar mass fraction for each IMF as well as that
from lensing and kinematics.

We find that, for the range of assumed luminous profiles and
anisotropies, the stellar mass
fraction from lensing agrees remarkably well
with that from stellar population modeling assuming a Salpeter IMF
(Fig.~\ref{fig:imf}). This result further supports the results found
by Treu et al. (2010) that the IMF of massive early-type lens galaxies
is well-matched by an evolved Salpeter IMF.
They found a tentative trend of the IMF slope with galaxy velocity dispersion, with a 'light' IMF such as a Chabrier IMF is inappropriate for systems with  $\sigma \geq 200\kms$.
This trend may imply a non-universal IMF, perhaps dependent on metallicity, age, or abundance ratios of the stellar populations. Alternatively, it may imply non-universal dark matter halos with inner density slope increasing with velocity dispersion.
While the degeneracy between the two interpretations cannot be broken without additional information, Treu et al. data imply that massive early-type galaxies cannot have both a universal IMF and universal dark matter halos. This is confirmed by the expanded analysis of Auger et al. 2010, who added weak-lensing data to constrain the halo model, finding that the data for massive galaxies are inconsistent with a Chabrier universal IMF for a range of realistic halo profiles, including various recipes to account for baryonic effects. 

\section{BOTTOM-HEAVY IMFs}

Recently, van Dokkum \& Conroy (2010) have suggested the presence of a
large number of stars with masses $\leq 0.3M_{\odot}$ in the central
regions of early-type galaxies.  By measuring the strength of the NaI
doublet and the Wing-Ford molecular FeH band in the spectra of eight
of the most luminous and massives early-type galaxies in the Virgo and
Coma clusters, they infer that the IMF for these systems might even be steeper
than Salpeter, with a slope as steep as $\alpha=-3$. They also test even 
more `bottom-heavy' with $\alpha=-3.5$ and `bottom-light' (dwarf deficient) IMF but they find the best fit between stellar population synthesis models and spectrum around the NaI doublet with the  $\alpha=-3$ IMFs, although the uncertainties are large and Salpeter cannot be excluded with high-confidence..  

Here, we investigate this claim by assuming a broken power-law IMF,
with the Salpeter slope in the high-mass regime that dominates the
SDSS magnitides (i.e. changes in the IMF below this break do not
affect the stellar-population analysis carried out above) and a
steeper profile in the low-mass range.  We test three different values
of the break point in the mass function: $m_{break}=0.3, 0.5 {\rm
  ~and~} 0.7\,{M}_{\odot}$, respectively.  Changing the IMF for stars
with $M \leq m_{break}$ does not change the SDSS colors because $\ga
90$ per cent of the light in the spectral region we studied here is
coming from stars with $M\geq 0.7\,{M}_{\odot}$. On the other hand, it
dramatically changes the total stellar mass of the system, because
stars with masses of $0.1 - 0.3 {M}_{\odot}$ can contribute at least
60 per cent of the stellar mass for these bottom-heavy IMFs.

We calculate the change in total stellar mass arising from the change
in the slope of the IMF:
\begin{equation}
\Delta M=\int_{0.1M_{\odot}}^{m_{\rm break}} \left[\left . \dfrac{dN}{dm}\right|_{\rm IMF}-\left . \dfrac{dN}{dm}\right|_{\rm Salp} \right] \, m \, dm,
\end{equation}

where

\begin{equation}
\left . \dfrac{dN}{dm}\right|_{\rm IMF}=\left\{
\begin{array}{l l}
\left(\frac{m}{m_{\rm break}}\right)^{\eta} & \quad \text{if~ $m_{\rm break} < m \le 100\,{\rm M}_{\odot}$}\\
\\
\left(\frac{m}{m_{\rm break}}\right)^{\alpha}  & \quad \text{if~ $0.1\,{\rm M}_{\odot} < m\leq m_{\rm break}$}\\
\end{array}\right .,
\end{equation}
with $\eta=-2.35$ (Salpeter slope), $\alpha=-3$ for
the bottom-heavy IMFs suggested by van Dokkum \& Conroy (2010) or $\alpha=-3.5$ in the most extreme case.

Our results for the bottom-heavy stellar mass fraction are listed in
Table \ref{tab:stmass}. We also list the likelihood ratios between the
nominal isotropic Hernquist model lensing and dynamic stellar mass
fraction and the stellar mass fractions obtained from stellar
populations and colors for the different IMFs (equivalent to $\Delta
\chi^{2}$ if the distribution was Gaussian, which we assume as a first
approximation), comparing their maximum \emph{a poteriori} difference
with the no-difference hypothesis. Vertical arrows in
Figure~\ref{fig:imf} show the stellar mass fractions predicted by
stellar population synthesis models and SDSS color by these different
IMFs and the stellar mass fraction obtained with lensing and
dynamics. Using $m_{\rm break}=0.7M_{\odot}$ and a bottom-heavy IMF,
we find that inferred stellar mass fraction exceeds unity for
$\alpha=-3.5$, inconsistent with the lensing mass.  An extreme
`bottom-heavy' IMF with slope of $\alpha = -3.5$ is inconsistent at
the $>90$ per cent confidence level with the lensing and kinematic
results.  The $\alpha=-3$ model is only marginally consistent for
$m_{\rm break}=0.3$ but is also excluded at the $>90$ per cent
confidence level for $m_{\rm break}=0.7$.  A Salpeter IMF gives the
best agreement with the lensing and kinematics.
It is important to mention that we do not include any possible effects of large-scale structure line-of-sight contamination (e.g. from a group elongated along the line of sight), that would decrease the total mass assumed here. 
Recalling Treu et al. (2009) and Guimaraes and Laerte Sodr\'{e} Jr (2011), SLACS lenses are shown to be unbiased sample in relation to a random LOS, despite the fact that the lenses are elliptical galaxies which are often found in dense regions.
Moreover, other possible explanations, such as
a top heavy IMF with more black holes and neutron stars remnants, are still possible.
As discussed in van Dokkum (2008), top-heavy IMFs have fewer low-mass stars  than a standard Salpeter IMF
but many more high-mass stars. 
Nevertheless, it is important to clarify that since our method only infers the overall $M/L$, 
we cannot distinguish a Salpeter IMF from a bottom-light IMF like a Chabrier IMF due to the presence of remnants (Treu et al. 2010 and Auger et al. 2010). 
    
\section{Summary \& Conclusions}

In this paper we present the first results from a new spectroscopic
survey of massive early-type lens galaxies: The X-Shooter Lens Survey
(XLENS). The combination high-fidelity UVIS-IR spectroscopy from the
X-Shooter instrument on the VLT, with the strong gravitational lensing
mass determination has enabled us to conduct an in-depth study of the
internal structure of the luminous elliptical galaxy SDSS J1148+1930
at $z=0.444$. We find the following:

\begin{enumerate}

\item The luminosity-weighted stellar velocity dispersion is $\langle
  \sigma_{*}\rangle(\la R_{\rm eff})=352\pm10\pm16\,\kms$, more accurate
  and considerably lower than a previously published value of
  $\sim450\,\kms$.

\item A single-component (stellar plus dark) mass model of the lens
  galaxy yields a logarithmic total-density slope of
  $\gamma'=1.72^{+0.05}_{-0.06}$ (68 per cent CL; $\rho_{\rm tot}
  \propto r^{-\gamma'}$).

\item The projected stellar mass fraction, derived solely from the
  lensing and dynamical data, is $f_{*}(<R_{\rm E}) =
  0.19^{+0.04}_{-0.09}$ (68 per cent CL) inside the Einstein radius
  for a Hernquist profile and no anisotropy. The dark-matter fraction
  inside the effective radius $f_{\rm DM}(<R_{\rm eff}) =
  0.60^{+0.15}_{-0.06} \pm 0.1$ (68 per cent CL), where the latter
  error is systematic.

\item Based on the SDSS colors, we find $f_{*, \rm Salp}(<R_{\rm
  E})=0.17 \pm 0.06$ for a Salpeter IMF and $f_{*,\rm Chab}(<R_{\rm
  E})=0.07 \pm 0.018$ for a Chabrier IMF. 
  A Salpeter IMF gives the best agreement between lensing and dynamics 
  constraints on the stellar mass fraction, therefore it is preferred to a Chabrier IMF.
  Dwarf-rich IMFs with $\alpha\ge 3$ (with
  $dN/dM \propto M^{-\alpha}$) in the lower
  mass range of $0.1$-$0.7\,M_{\odot}$,  -- such as those recently suggested by van Dokkum \&
  Conroy (2010) for massive early-type galaxies ($\alpha=-3$) -- are excluded at the $>90$
  per cent CL and in some cases ($\alpha=-3.5$) violate the total lensing-derived
  mass limit.

\end{enumerate}
This massive early-type galaxy lies at the extreme end of the trend
found by, e.g., Auger et al.\ (2010) that the dark matter fraction
within the effective radii is a monotonically increasing function of
galaxy mass. In fact SDSS J1148+1930 is already dark-matter dominated
within that region. 
We find that a Salpeter IMF agrees best with the total stellar mass
derived from lensing and stellar kinematics as well as with its SDSS
colors. As in Treu et al.\ (2010) and Grillo \& Gobat (2010), a Salpeter IMF appears to be the best option
for very massive early-type galaxies. Although slightly more massive IMFs cannot be excluded given the typical uncertainties.
A bottom-light IMF such as a Chabrier (or Kroupa) IMF agrees only
marginally and we exclude a steep `dwarf-rich' IMF with $\alpha=-3.5$
at $>90$ per cent CL. 
Somewhat shallower IMFs with $\alpha\approx -3.0$, as 
suggested by van Dokkum \& Conroy (2010), are
marginally acceptable.  We conclude that our data are {\sl fully}
consistent with SDSS J1148+1930 being a massive early-type galaxy that
is dark-matter dominated inside its effective radius and having a
Salpeter IMF. No strong evidence for an even more bottom-heavy IMF is found,
consistent with previous results (Treu et al.\ 2010), although uncertainties are still large..

Further studies are required to break the degeneracy between the
central dark-matter fraction and distribution and the stellar IMF. In
forthcoming papers of the XLENS survey, we will extend the study to
more massive systems at $z\ga 0.5$ and also a sub-sample of SLACS
lenses at $z\la 0.5$. Observations are ongoing. In those papers, we
will perform more detailed stellar population analyses using the full
UV-optical-NIR spectrum and obtain independent contraints on the the physical
parameters that may correlate with IMF normalization (i.e., age and
metallicity) or that may be the cause of the correlation between
dark-matter content and velocity dispersion.

\section*{Acknowledgments}
The authors thank the referee for providing constructive comments 
and help in improving the contents of this paper.\\
The use of the Penalized Pixel Fitting developed by Cappellari \&
Emsellem and of the GALAXEV Software by Bruzual \& Charlot are
gratefully acknowledged. Data were reduced using EsoRex and XSH
pipeline by ESO Data Flow System Group. We thank Prof. P. Groot for
instruction on its use. C.S. acknowledges support from an Ubbo Emmius
Fellowship. L.V.E.K. is supported in part by an NWO-VIDI program
subsidy (project number 639.042.505). 
T.T. acknowledges support from a Packard Research Fellowship.

\label{lastpage}
\end{document}